%% file: _main.tex
\documentclass[10pt, conference, letterpaper]{IEEEtran}
\IEEEoverridecommandlockouts

\usepackage[table,xcdraw]{xcolor}
\usepackage{eso-pic}
\usepackage{graphicx}
\usepackage{booktabs}
\usepackage{array}

\usepackage{mathtools}

 \usepackage[absolute]{textpos}
 
\input{_preamble}

\begin{document}

 \begin{textblock}{5}(9.8,0.8) (Accepted in COINS'25 Special Session) \end{textblock}

\bstctlcite{IEEEexample:BSTcontrol}
\IEEEtriggercmd{\balance}
\IEEEtriggeratref{25}

\title{%
    Towards LLM-based Root Cause Analysis of Hardware Design Failures%
}

\author{%
\IEEEauthorblockN{Siyu Qiu\IEEEauthorrefmark{1}, Muzhi Wang\IEEEauthorrefmark{1}, Raheel Afsharmazayejani\IEEEauthorrefmark{2}, Mohammad Moradi Shahmiri\IEEEauthorrefmark{2}\\ Benjamin Tan\IEEEauthorrefmark{2} and Hammond Pearce\IEEEauthorrefmark{1}}

\IEEEauthorblockA{%
\textit{\IEEEauthorrefmark{1}School of Computer Science and Engineering}, \textit{UNSW Sydney}}

\IEEEauthorblockA{%
\IEEEauthorrefmark{2}\textit{Department of Electrical and Software Engineering}, 
\textit{University of Calgary}}

\textit{benjamin.tan1@ucalgary.ca, hammond.pearce@unsw.edu.au}

}

\maketitle

\begin{abstract}
With advances in large language models (LLMs), new opportunities have emerged to develop tools that support the digital hardware design process. In this work, we explore how LLMs can assist with explaining the root cause of design issues and bugs that are revealed during synthesis and simulation, a necessary milestone on the pathway towards widespread use of LLMs in the hardware design process and for hardware security analysis. We find promising results: for our corpus of 34 different buggy scenarios, OpenAI's o3-mini reasoning model reached a correct determination 100\% of the time under pass@5 scoring, with other state of the art models and configurations usually achieving more than 80\% performance and more than 90\% when assisted with retrieval-augmented generation.

\end{abstract}
\begin{IEEEkeywords}
    EDA, root cause analysis, error explanation, LLM %
\end{IEEEkeywords}

\input{sec/01intro}

\input{sec/02background}

\input{sec/04method}

\input{sec/05results}

\input{sec/06Discussion}
\input{sec/07Conclusions}

\bibliographystyle{IEEEtran}
\bibliography{IEEEabrv, ref/trefs, ref/clean}

\end{document}

%% file: _preamble.tex
\usepackage{cite}%
\usepackage{amsmath,amssymb,amsfonts}
\usepackage{comment}
\usepackage{algorithmic}
\usepackage{graphicx}
\usepackage{textcomp}

\usepackage[all=normal, paragraphs=tight]{savetrees}

\usepackage{enumitem}

\usepackage{listings}

\usepackage{booktabs}  
\usepackage{csquotes}
\usepackage{multirow}
\usepackage{array}
\newcolumntype{L}[1]{>{\raggedright\let\newline\\\arraybackslash\hspace{0pt}}m{#1}}
\newcolumntype{C}[1]{>{\centering\let\newline\\\arraybackslash\hspace{0pt}}m{#1}}
\newcolumntype{R}[1]{>{\raggedleft\let\newline\\\arraybackslash\hspace{0pt}}m{#1}}
\usepackage{url}
\MakeOuterQuote{"}
\usepackage[caption=false,font=footnotesize,labelformat=simple]{subfig} %

\usepackage[bookmarks=false]{hyperref}

\definecolor{mygreen}{rgb}{0,0.6,0}
\definecolor{mygray}{rgb}{0.5,0.5,0.5}
\definecolor{mymauve}{rgb}{0.58,0,0.82}

\makeatletter
\newcommand\notsotiny{\@setfontsize\notsotiny\@vipt\@viipt}
\makeatother

\usepackage{acronym}
\input{z_acroynms}

%% file: z_acroynms.tex
\acrodef{IP}[IP]{intellectual property block}
\acrodef{SoC}[SoC]{System-on-Chip}
\acrodef{IC}[IC]{integrated circuit}
\acrodef{eFPGA}[eFPGA]{embedded field programmable gate array}
\acrodef{LLM}[LLM]{large language model}
\acrodef{EDA}[EDA]{electronic design automation}

%% file: sec/01intro.tex
\section{Introduction\label{intro}}

Encountering bugs, glitches, and faults is a normal part of the digital hardware design lifecycle. To ensure they are completely removed and repaired is a time-consuming process requiring a deep understanding of both the technical cause of the issue as well as any impacts on the broader hardware system -- particularly as any missed repair may have severe downstream functional and/or security consequences~\cite{dessouky_hardfails_2019} (if the bug is of an exploitable nature). 
However, as digital hardware grows in complexity, so do the frequency and nature of the bugs themselves. We are therefore motivated to seek new tools and techniques that can aid in the digital design process, simplifying this complex task.

Transformer-based \acp{LLM}%
have recently demonstrated tremendous performance at various tasks in the electronic design automation (EDA) flow, including fixing syntax~\cite{tsai_rtlfixer_2024},
writing HDL to pass design test suites~\cite{liu_invited_2023,pinckney_revisiting_2025} and
automatically integrating tool feedback to improve such designs~\cite{blocklove_automatically_2025}, and aid in the design of relatively complex hardware, including AI accelerators~\cite{fu_gpt4aigchip_2023} and microcontrollers~\cite{blocklove_chip-chat_2023}. 

Given this performance, we therefore muse: are LLMs also capable of understanding hardware faults and determining the root causes of design failures such that they may produce reasonable fault explanations? This is particularly pertinent as prior work has suggested they may be used to explain synthesis error messages~\cite{qiu_llm-aided_2024}.
Further, while most recent research on LLMs and hardware proposes and aims to improve LLMs that operate autonomously (i.e., with minimal human intervention), we envisage a future where LLMs and human engineers work \textbf{side-by-side} to efficiently design hardware. As such, investigations such as this, where we explore how an LLM might aid a design engineer, are useful to understand how they might be best used in the human-LLM design loop.

To determine the answer to our question, in this work, we explore a corpus of buggy scenarios (19 synthesis-time and 15 simulation run-time errors) with four state-of-the-art LLMs (\texttt{gpt-3.5-turbo}, \texttt{GPT-4o}, \texttt{o3-mini}, and \texttt{DeepSeek R1}). We prompt each LLM to identify the root cause of each error and how to mitigate it, then evaluate their answers manually. In line with current best practice in LLM prompting, we repeat the experiment using Retrieval-Augmented Generation (RAG). This process is shown in~\autoref{fig:methodology} and expanded upon in~\autoref{sec:method}. We frame our investigation around the following research questions:

\begin{figure}[tbp]
    \centering
    \includegraphics[width=0.9\linewidth]{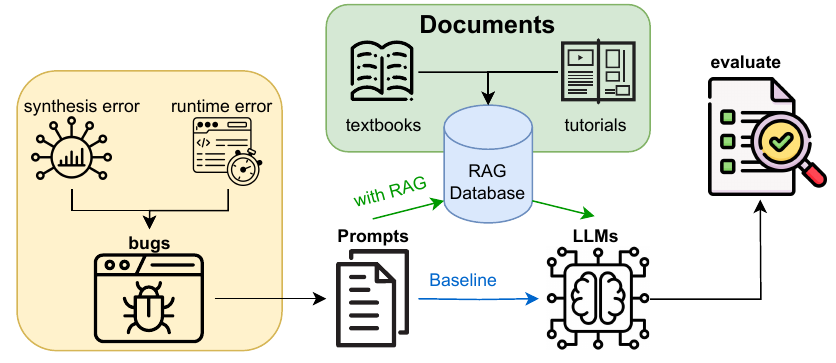}
    \vspace{-1em}
    \caption{Overall experimentation method\label{fig:methodology}}
    \vspace{-2em}
\end{figure}

\IEEEpubidadjcol

\begin{itemize}[leftmargin=8mm]
    \item[\textbf{RQ1}] Can LLMs accurately and meaningfully explain the root cause of digital hardware design failures occurring at synthesis time?
    \item[\textbf{RQ2}] Likewise, can LLMs accurately and meaningfully explain root causes of failures from simulation runtimes?
    \item[\textbf{RQ3}] Can we improve their performance when using standard LLM prompt engineering techniques, i.e. Retrieval-Augmented Generation (RAG)?
\end{itemize}

\noindent The contributions of this work are as follows:
\begin{itemize}[leftmargin=*]
    \item A corpus of permissively-licensed open-source and annotated hardware projects containing synthesis-time and runtime design failures, reusable for other research tasks outside the scope of this work (e.g. automated bug repair), made available at \url{https://github.com/llm4hw/root-cause-bench}.
    \item The first systematic exploration of hardware design failure root cause analysis by LLMs. 
    \item Insights into future directions for LLM use and evaluation.
\end{itemize}

%% file: sec/02background.tex
\section{Background and Motivation}
\label{sec:bg}

Recent developments in \acp{LLM} have motivated research on their utility in both hardware and software-related development tasks. 
Early work included the use of \acp{LLM} for creating simple Verilog designs~\cite{pearce_dave_2020}; recent work has examined LLMs creating digital hardware autonomously (e.g.,~\cite{blocklove_automatically_2025})
Related prior work involves the use of \acp{LLM} in tasks such as writing SystemVerilog assertions (e.g.,~\cite{maddala_laag-rv_2024}) and creating test benches (e.g.,\cite{qiu_autobench_2024}).
Security-related LLM work has included security assertion generation~\cite{kande_security_2024}, bug fixing~\cite{ahmad_fixing_2023}, and bug detection~\cite{ahmad_flag_2025}.
Broadly speaking, prior work focuses on \acp{LLM} tasked with generating structured technical outputs (i.e., code). 

To complement these prior efforts, we focus instead on exploring LLMs and their capabilities in \textit{analyzing and explaining} hardware artifacts (specifically, hardware errors), with outputs in plain language. 
Essentially, we are interested in seeing how LLMs can help \textit{support} human designers (instead of only automating parts of the EDA flow), an area that has yet to be thoroughly explored in the hardware domain.

There are several motivations for exploring this space. 
For instance, by using LLMs to \textit{explain} rather than opaquely and autonomously enact changes, we retain a human-in-the-loop with ultimate responsibility for the design. 
This also presents an opportunity to upskill human designers by learning from LLM assistants. 
Research in the software space currently leads hardware; Chapagain et al. \cite{ferreira_mello_technology_2024} investigated the capabilities of LLMs to generate line-by-line explanations of programming code, and Taylor et al. \cite{taylor_dcc_2024} considered the combination of LLMs and debugging compilers -- these works suggest that LLMs can help novices enhance their understanding by improving error messages, in line with the findings made by Karvelas et al.~\cite{karvelas_effects_2020}.

In hardware, prior work~\cite{qiu_llm-aided_2024} %
showed that out-of-the-box LLMs can enhance the readability of EDA tool error messages.
This work extends previous work by examining errors that arise in synthesis and issues that present during simulation. 
We task an LLM with explaining what has gone wrong. 
This study serves as a starting point that can lead to future studies of LLMs and their ability to handle other parts of the EDA flow, including security analyses.  

Investigating the use of LLMs in explaining presents a methodological challenge; there do not yet exist easily automated evaluation approaches for the ``quality'' of explanations of hardware flaws. 
Thus, we adopt the manual evaluation approach of prior work~\cite{qiu_llm-aided_2024} where we manually grade outputs on binary (yes/no) criteria: \textit{concept accuracy}, \textit{no inaccuracies} (factuality), \textit{relevance}, and \textit{correct \& complete}. Of these, the most important metric is \textbf{correct \& complete}, which means that the explanation is factual, explains the root cause, and describes an approach to fix the problem.

%% file: sec/04method.tex
\section{Evaluation Method}
\label{sec:method}

\subsection{Overview}

To answer our RQs, we develop an approach that uses \acp{LLM} to assist in the debugging of hardware design errors. 
As shown in \autoref{fig:methodology}, the process starts with synthesis and runtime error data generated in an \ac{EDA} tool, for example, AMD's Vivado.
Each error message (from the tool or simulation transcript) is combined with a template to produce a prompt for the \ac{LLM}. 
To evaluate the \ac{LLM}, we manually scored the generated responses as in prior work~\cite{qiu_llm-aided_2024}.

\subsection{Bug Dataset Construction}
To evaluate the ability of the \ac{LLM} to identify errors, we constructed a dataset of manually-labelled hardware design errors in Vivado. 
The dataset is divided into two categories:

\begin{itemize}

\item \textbf{Synthesis errors:} Errors triggered during the synthesis phase, usually due to incorrect syntax, type mismatches, illegal assignments, or unsupported HDL constructs.

\item \textbf{Runtime errors:} Errors that occur during the simulation phase, usually due to problems with the testbench or simulation behaviour, which affect functional correctness.

\end{itemize}

\noindent Each error entry in the data set contains:
\begin{itemize}
  \item HDL source code (Verilog or VHDL)
  \item Project files, including .xpr and .xdc (constraint) files
  \item Minimal testbench (for runtime errors only)
\end{itemize}

\noindent The entire dataset contains 34 errors: 19 synthesis errors and 15 runtime errors. We diversify both categories by covering a wide range of representative bug types. Detailed listings of each error are provided in~\autoref{tab:synth_bugs} (synthesis errors) and~\autoref{tab:runtime_bugs} (runtime errors).
These errors are mostly simple design failures where the runtime bugs represent flawed logic that, while we do not make explicit, could also apply to security.

\begin{table*}[]
\centering
\renewcommand{\arraystretch}{0.8}
\caption{Summary of Synthesis Error Bugs with Descriptions -- These 19 bugs are the Vivado-only subset from \cite{qiu_llm-aided_2024}}
\label{tab:synth_bugs}
\scriptsize
\begin{tabular}{|C{0.5cm}L{2cm}L{1cm}L{3.8cm}||C{0.5cm}L{2cm}L{1cm}L{3.8cm}|}
\hline
\textbf{Bug} & \textbf{Error Type} & \textbf{Lang.} & \textbf{Error Description} & \textbf{Bug} & \textbf{Error Type} & \textbf{Lang.} & \textbf{Error Description} \\ \hline
V1 & Syntax error & VHDL & Missing semicolon & V12 & Signal bit error & VHDL & Mismatch between a \textbackslash{}texttt\{std\_logic\} type and a string literal \\ \hline
V2 & Type error & VHDL & Can't add \texttt\{std\_logic\_vectors\} & V13 & Syntax error & Verilog & Missing semicolon \\ \hline
V3 & Compilation error & VHDL & Can't write to an input ports object & V14 & Semantic error & Verilog & Using an undeclared variable or signal \\ \hline
V4 & Width mismatch & VHDL & Mismatch in the size of two \texttt\{std\_logic\_vectors\} & V15 & Wire and Reg error & Verilog & Assigning a value to a \texttt\{wire\} using non-blocking assignments \\ \hline
V5 & Type conversion & VHDL & Can't perform two operations simultaneously in one line & V16 & Blocking and non-blocking & Verilog & Mixing blocking and non-blocking assignments to the same variable \\ \hline
V6 & Signal and variable & VHDL & Declaring a variable outside of a subprogram or process & V18 & Port error & Verilog & Connecting a port that does not exist \\ \hline
V7 & Concurrent and sequential error & VHDL & Having both \texttt\{wait\} and a sensitivity list in the same process & V19 & Binary error & Verilog & Using an illegal character in a binary number representation \\ \hline
V8 & Semantic error & VHDL & Using a signal or variable that has not been declared & V20 & Infinite combinational loop & Verilog & Having an infinite combinational loop that cannot be resolved \\ \hline
V9 & Signal readability error & VHDL & Attempting to read from an object with the mode \texttt\{out\} & V21 & Double-edge error & Verilog & Mismatch between operands used in the condition of an \texttt\{always\} block \\ \hline
V11 & Case error & VHDL & Missing certain choices in a \texttt\{case\} statement & \cellcolor[HTML]{9B9B9B} & \cellcolor[HTML]{9B9B9B} & \cellcolor[HTML]{9B9B9B} & \cellcolor[HTML]{9B9B9B} \\ \hline
\end{tabular}%
\end{table*}

\begin{table*}[ht]
\centering
\caption{Summary of runtime Bugs with Descriptions and Error Messages (in Verilog only)}
\label{tab:runtime_bugs}
\renewcommand{\arraystretch}{1.1}
\scriptsize
\begin{tabular}{|c l p{7.3cm} p{6cm}|}
\hline
\textbf{Bug} & \textbf{Error Type}  & \textbf{Error Description} & \textbf{Error Message} \\
\hline
R1  & Combinational Logic         & Incorrect OR and AND logic combination & ERROR: Test Case 1 failed. p1y = 0 (Expected: 1) \\
\hline
R2  & Combinational Logic         & AND operation with constant 0 - always results in 0 & ERROR: Test Case 4 failed. q = 0 (Expected: 1) \\
\hline
R3  & Carry Logic Error           & Incorrect carry logic in full adder implementation & ERROR: Test Case 3 failed. o\_sum = 0, o\_carry = 0 (Expected: 0, 1) \\
\hline
R4  & State Machine Conflict      & Duplicate case statement causing unexpected output & ERROR: Test Case 1 failed. f = 0 (Expected: 1) \\
\hline
R5  & Sequential Logic Error      & Incorrect sequential check for first smaller value & ERROR: Test Case 3 failed. min = 7 (Expected: 8) \\
\hline
R6  & FSM State Transition Error  & Incorrect duration logic for Ped light with longTimeHold & ERROR: Test Case 6 failed. outResult = 101 (Expected $>$ 110) \\
\hline
R7  & FSM State Transition Error  & Incorrect state transition from Ped to Red instead of Green & ERROR: Expected Red (00), but got 11 \\
\hline
R8  & FSM State Transition Error  & Incorrect transition from Ped to Green when pedControl is held & pedControl 2 - ERROR: Expected Green (01), but got 11 \\
\hline
R9  & FSM State Transition Error  & Incorrect Morse code generation for letter 'O' & ERROR: Morse for O not correct. \\
\hline
R10 & Shift Register Logic Error  & Incorrect shifting logic in 4-bit shift register & ERROR: Expected 0001 got 1111 \\
\hline
R11 & Input Validation Error      & FSM does not correctly handle invalid numeric input outside 0--9 range & ERROR: Invalid input should not be valid. \\
\hline
R12 & Bit Width Mismatch          & Comparator uses 5-bit constant to compare with 4-bit input & ERROR: Expected result = 1 when a = b \\
\hline
R13 & Wiring Error                & Incorrect carry chain wiring and final carry output in 4-bit full adder & ERROR: a = 3, b = 5, cin = 0 $\rightarrow$ Expected 01000, got 10010 \\
\hline
R14 & Sensitivity List Error      & FSM uses posedge reset instead of posedge clk, preventing state transitions & Test 1 failed: Expected state = 3, output\_data = 3, valid = 1, got 0 0 0 \\
\hline
R15 & Structural Error            & Ring oscillator failed to toggle due to lack of startup trigger or improper feedback loop & ERROR: Ring oscillator output (OUT) never toggled. Likely feedback wiring issue. \\
\hline
\end{tabular}
\end{table*}

\subsection{Prompt Design and Models}

\acp{LLM} are used to generate natural language responses given a prompt. 
In this study, we evaluated three OpenAI models: \texttt{gpt-3.5-turbo}, \texttt{gpt-4o}, and \texttt{o3-mini}, as well as the \texttt{Deepseek R1} model (evaluation date: 2025-03 to 2025-05).

Each model requires a system prompt to establish its role, followed by a user prompt with task-specific content.

To ensure consistency and fairness in the evaluation, we adopted a unified prompting strategy for both synthesis errors and runtime errors. We base our prompting on previous studies~\cite{qiu_llm-aided_2024}, but with an additional prompt input for runtime errors as shown in~\autoref{fig:prompts}:
\begin{figure}[t]
    \centering
    \begin{lstlisting}
Note: This error occurred during simulation of the testbench and the following code files. This is an error message generated by the testbench itself. You should identify the root cause of the problem and explain the fix.
    \end{lstlisting}
    \vspace{-2mm}
    \caption{Our addition to the LLM prompts from \cite{qiu_llm-aided_2024}}
    \label{fig:prompts}
    \vspace{-1em}
\end{figure}
This supplementary information explicitly instructs the model to explain simulation-specific errors and provide corrective reasoning.

For synthesis errors, hints follow the same structure but do not include a testbench. Instead, they contain the specific line number where the error occurred. All hints are presented in plain text. We intend for the model to explain the root cause and suggest a reasonable fix.

For each model, we generate 5 responses for each error using the same prompt. There are 15 runtime errors and 19 synthesis errors in the dataset, for a total of
5×(15+19)=170 responses; the authors grade each manually. This sampling strategy supports random evaluation metrics (e.g., pass@k) and measuring intramodel debugging quality differences.

\subsection{Retrieval-Augmented Generation}
Based on current LLM practices, we augmented the \ac{LLM}’s external retrieval capabilities with a custom RAG system, with the hope of improving diagnostic accuracy. %
For each input, the system retrieves relevant documents and adds them to the prompt, allowing the model to reason with task-specific prior knowledge.
The RAG database contains authoritative content on VHDL and Verilog, including \textbf{textbook content} (excerpts from digital logic, Verilog/VHDL books, etc., with formal definitions, common pitfalls, and explanations of design and test practices) and \textbf{tutorial/practical guides} (e.g., online labs, Vivado tool guides, and other ``application note'' docs).

These documents are preprocessed into blocks, embedded using OpenAI's \texttt{text-embedding-3-small} model, and indexed by ChromaDB for semantic retrieval. 
For each error query, we extract the code and error message as query vectors and retrieve the top 6 most relevant pieces of information.
The retrieved content is inserted into the prompt and combined with the AI model to improve accuracy.

\subsection{Pass@k Computation}
To evaluate the model's performance in the random case, we use the Pass@k metric~\cite{chen2021evaluating}, which is commonly employed in code generation and multi-sample evaluation settings.

Formally, Pass@k is computed as:

\begin{equation}
\operatorname{pass@}k \coloneqq \mathbb{E}_{\text{problems}} \left[ 1 - \frac{\binom{n - c}{k}}{\binom{n}{k}} \right]
\end{equation}

\noindent where:
\begin{itemize}
  \item $n$: total number of generated solutions;
  \item $c$: number of correct solutions among the $n$;
  \item $k$: number of samples drawn uniformly without replacement;
  \item $\binom{n - c}{k}$: number of $k$-combinations from incorrect solutions;
  \item $\binom{n}{k}$: number of all possible $k$-combinations;
  \item $\mathbb{E}_{\text{problems}}$: expectation over all evaluated problems.
\end{itemize}

\noindent At the same time, if n=k, the estimator based on expected sampling \cite{chen2021evaluating} is simplified to a binary indicator to indicate whether there is at least one correct solution, as follows:

\begin{equation}
\text{Pass@}k = \frac{n_{\text{pass}}}{n_{\text{total}}}
\end{equation}

\noindent where:
\begin{itemize}
    \item $k$: the number of sampled predictions per problem;
    \item $n_{\text{pass}}$: the number of problems among all test problems that have at least one candidate solution that passes all test cases;
    \item $n_{\text{total}}$: the total number of problems evaluated.
\end{itemize}

\noindent In our setup, for each error, the model generates 5 different responses. If at least one of these 5 responses is rated correct, the model passes the corresponding metric. This metric reflects the model's best-case performance under $k$ attempts.

\subsection{Avg@k Computation}

This is much a simpler metric which reflects the total number of model answers which were correct divided by the number of attempts; and when aggregating across problems it is the summation of the aforementioned averages divided by the total number of problems evaluated. Where Pass@$k$ gives you a probability for any one of a set of $k$ responses to be correct, Avg@$k$ provides a way to estimate whether or not any one response is correct.

\subsection{Explanation Quality}

To illustrate the qualitative differences between the model outputs, we provide two example explanations for the same synthetic runtime error (Error 3) in Figure~\ref{fig:explanation} (recall from Section~\ref{sec:bg} that we are using the explanation quality metrics from \cite{qiu_llm-aided_2024}).

The `good' explanation generated by \texttt{DeepSeek R1} correctly points out that the error is caused by using an AND gate instead of an OR gate in the carry output logic. It provides a clear explanation and rationale, and this output receives full scores on all dimensions.
The `bad' explanation generated by \texttt{gpt-3.5-turbo} focuses on signal declaration and module instantiation, both of which are irrelevant to the actual error. The response contains factual errors and fails to identify the root cause.

\begin{figure}[t]
    \centering
    \begin{lstlisting} 
The bug is in the calculation of the final carry output (`o_carry`). 

**Why it's occurring:**  
The full adder's carry should be the logical **OR** of the two half-adder carry outputs (`w_carry1` and `w_carry2`), as a carry occurs if either of the two additions (first: `i_a + i_b`, second: intermediate sum + `i_cin`) produces a carry. However, the code erroneously uses an **AND** operation (`w_carry1 & w_carry2`) to combine the carries. This causes `o_carry` to only be asserted when both half-adders generate a carry, which is incorrect for standard full-adder behavior.

**How to fix it:**  
Replace the **AND** operator with an **OR** operator in the assignment of `o_carry` to correctly propagate a carry when either half-adder produces a carry.

    \end{lstlisting}
    \vspace{-2mm}
    {\footnotesize (a) `Good' explanation---Concept accurate: \textbf{Yes}; Any inaccuracies: \textbf{No}; Relevant: \textbf{Yes}; Correct \& complete: \textbf{Yes}, Solution is provided: \textbf{No}}
    \begin{lstlisting}
The bug in the provided code is in the instantiation of the first half adder `h1` in the `full_half_add_1bit` module. The inputs `i_a` and `i_b` are directly connected as inputs to the `half_adder h1` module, which results in `i_b` being driven by two different sources, violating the Verilog single-source rule.

This violation causes incorrect values to propagate through the design, leading to incorrect results in the test cases that involve an `i_b` input value of 1. Specifically, Test Case 3 is affected because `i_b` is being affected by the result of `h1`.

To fix this bug, the output `w_sum1` from the first half adder `h1` should be used as the input `h_b` for the second half adder `h2`, rather than directly using `i_b`. By correcting the input connection in the instantiation of `h2`, the issue of multiple drivers for `i_b` and hence wrong test case outputs can be resolved.
    \end{lstlisting}
    \vspace{-2mm}
    {\footnotesize (b) `Bad' explanation---Concept accurate: \textbf{No}; Any inaccuracies: \textbf{Yes}; Relevant: \textbf{No}; Correct \& complete: \textbf{No}, Solution is provided: \textbf{No}}
    \vspace{-1mm}
    \caption{Example of `good' and `bad' error explanations for Bug 3 (Figure~\ref{tab:runtime_bugs}) generated by \texttt{DeepSeek R1} and \texttt{gpt-3.5-turbo}. Each bug is presented with graded metrics.}
    \label{fig:explanation}
    \vspace{-4mm}
\end{figure}

%% file: sec/05results.tex
\section{Results}

\subsection{Top-line performance}

\begin{figure}[t]
    \centering
    \includegraphics[width=0.8\columnwidth]{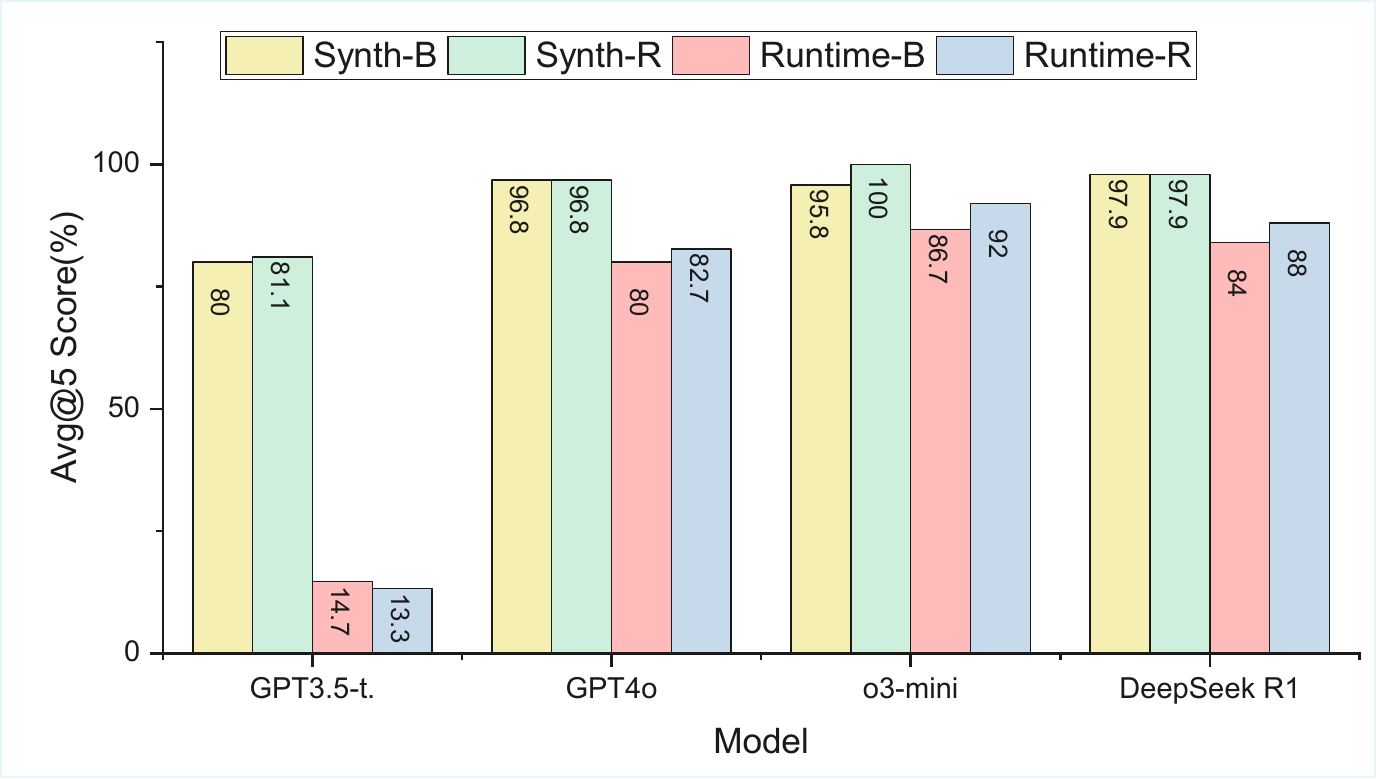}
    \vspace{-1em}
    \caption{Avg@5 Score (\%) across different models, under baseline (B) and RAG-enhanced (R) settings.}
    \vspace{-1em}
    \label{fig:avg5}
\end{figure}

\begin{figure}[t]
    \centering
    \includegraphics[width=0.8\columnwidth]{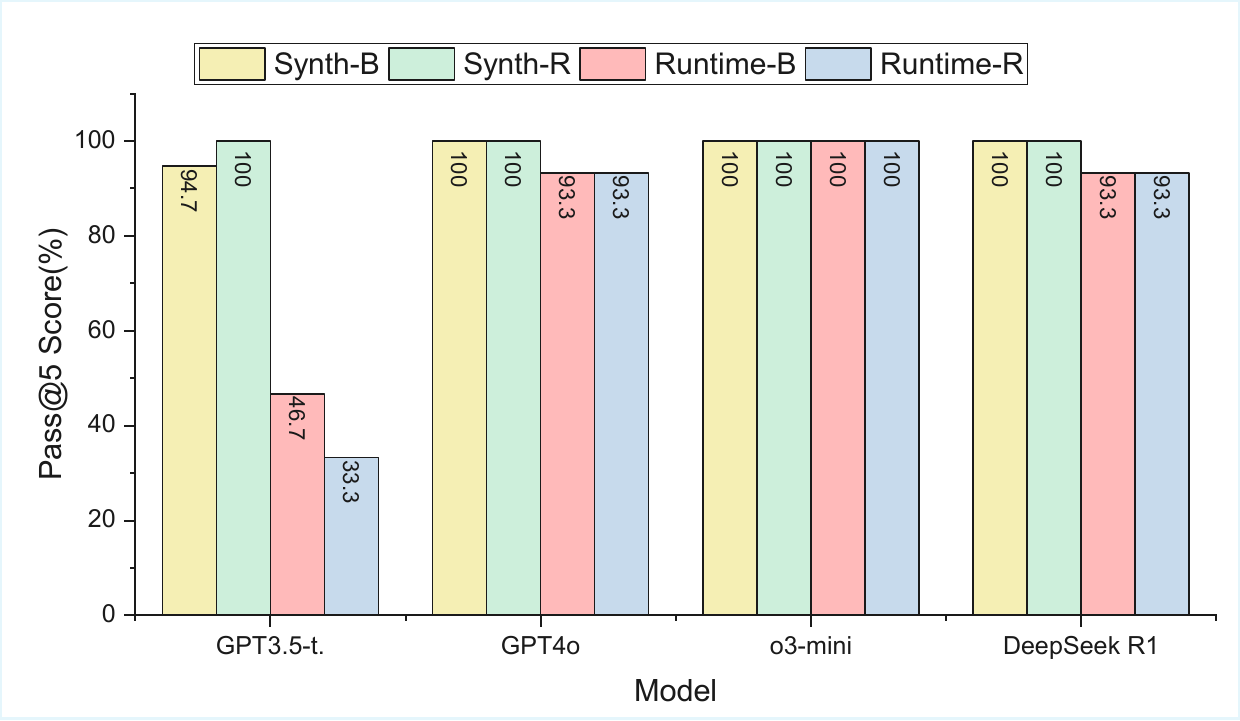}
    \vspace{-1em}
    \caption{Pass@5 Score (\%) for runtime and synthesis bugs. RAG improves runtime performance in most models except \texttt{gpt-3.5-turbo}.}
    \vspace{-1em}
    \label{fig:pass5}
\end{figure}

We compare the performance of each model in terms of runtime and synthetic error diagnosis under the baseline setting and the RAG-enhanced setting. The summary results are shown in ~\autoref{fig:avg5} and ~\ref{fig:pass5}, and the detailed metrics are listed in~\autoref{tab:reslut_combined_table}.

As can be seen, and particularly with RAG and with the three newer models \texttt{gpt-4o}, \texttt{o3-mini}, and \texttt{DeepSeek R1} the LLMs could accurately and meaningfully explain in a majority of cases the root cause of hardware design failures in both simulation and runtime scenarios (Ans. \textbf{RQ1}, \textbf{RQ2}.).

When exploring the performance with RAG, we see that it brings small to moderate improvements to most models (Ans. \textbf{RQ3}). Notably, \texttt{GPT-4o} and \texttt{o3-mini} benefit from RAG, with a higher Avg@5 and ``Correct \& Complete'' scores, indicating that the retrieved contextual information is effectively utilized. They perform most consistently across all conditions, maintaining high levels in both baseline and RAG tests, and with minimal differences between runtime and synthesis tasks.

\texttt{DeepSeek R1} also improves under RAG, with Avg@5 scores of over 88\% for concept accuracy, relevance, and correctness, as shown in Table~\ref{tab:reslut_combined_table}, indicating that its reasoning is positively influenced by the RAG database. 
While slightly inferior to GPT-4o in the baseline test, it significantly improves runtime performance in the RAG test (e.g., Avg@5 concept accuracy: 93.3\%) and achieves an average Avg@5 concept accuracy of 98.9\% in the synthesis-RAG. This shows that DeepSeek R1 has strong in-context learning capabilities. 

\texttt{gpt-3.5-turbo} is the only model that is hindered with RAG, dropping from 46.7\% to 33.3\% on Pass@5 and also reducing performance with Avg@5. This model is the weakest overall, likely due to its reduced base capability, knowledge, and size compared with its state-of-the-art peers. It is the weakest model overall, especially in the runtime RAG test, with low correctness and solution quality scores. Its sensitivity to context length and irrelevant retrieval makes it unsuitable for complex HDL debugging tasks.

In the synthesis setting, all models perform strongly on the baseline, with most reaching 100\% on Pass@5. RAG provides little additional improvement in this regard, which suggests that current large models have strong independent reasoning capabilities for synthesis error scenarios with clear rules and stable structures.

\subsection{Pass@5 and Avg@5}

Figures~\ref{fig:avg5} (Avg@5, showing us consistency) and~\ref{fig:pass5} (Pass@5, showing us potential) demonstrate that all LLMs have at least some potential utility when tasked with root cause analysis. \texttt{o3-mini} in particular had very strong results, made even better when augmented with RAG.
A good performance in both metrics, achieved by all models aside from \texttt{gpt-3.5-turbo} indicates that they not only have a chance to get the correct explanation but they also are stable and reliable at doing so.

\begin{table*}[t]
\centering
\caption{Comparison of LLM Performance across Task Types and Prompting Settings (Pass@5 and Avg@5 per model)}
\label{tab:reslut_combined_table}
\renewcommand{\arraystretch}{1.1}
\begin{tabular}{lcccccccc|}
\cline{2-9}
\multicolumn{1}{l|}{\textbf{}} & \multicolumn{2}{c|}{\textbf{GPT3.5-t.}} & \multicolumn{2}{c|}{\textbf{GPT4o}} & \multicolumn{2}{c|}{\textbf{o3-mini}} & \multicolumn{2}{c|}{\textbf{DeepSeek R1}} \\ \cline{2-9} 
\multicolumn{1}{l|}{\textbf{}} & Pass@5 & \multicolumn{1}{c|}{Avg@5} & Pass@5 & \multicolumn{1}{c|}{Avg@5} & Pass@5 & \multicolumn{1}{c|}{Avg@5} & Pass@5 & Avg@5 \\ \hline
\multicolumn{9}{|c|}{\textit{Runtime Baseline}} \\ \hline
\multicolumn{1}{|l|}{Concept accurate} & 73.3\% & \multicolumn{1}{c|}{44.0\%} & 100.0\% & \multicolumn{1}{c|}{90.7\%} & 100.0\% & \multicolumn{1}{c|}{94.7\%} & 93.3\% & 92.0\% \\
\multicolumn{1}{|l|}{No inaccuracies} & 66.7\% & \multicolumn{1}{c|}{33.3\%} & 100.0\% & \multicolumn{1}{c|}{88.0\%} & 100.0\% & \multicolumn{1}{c|}{93.3\%} & 93.3\% & 89.3\% \\
\multicolumn{1}{|l|}{Relevant} & 66.7\% & \multicolumn{1}{c|}{54.7\%} & 93.3\% & \multicolumn{1}{c|}{90.7\%} & 100.0\% & \multicolumn{1}{c|}{96.0\%} & 93.3\% & 92.0\% \\
\multicolumn{1}{|l|}{Correct \& complete} & 46.7\% & \multicolumn{1}{c|}{14.7\%} & 93.3\% & \multicolumn{1}{c|}{80.0\%} & \textbf{100.0\%} & \multicolumn{1}{c|}{\textbf{86.7\%}} & 93.3\% & 84.0\% \\ \hline
\multicolumn{9}{|c|}{\textit{Runtime RAG}} \\ \hline
\multicolumn{1}{|l|}{Concept accurate} & 46.7\% & \multicolumn{1}{c|}{28.0\%} & 100.0\% & \multicolumn{1}{c|}{93.3\%} & 100.0\% & \multicolumn{1}{c|}{93.3\%} & 93.3\% & 93.3\% \\
\multicolumn{1}{|l|}{No inaccuracies} & 40.0\% & \multicolumn{1}{c|}{17.3\%} & 100.0\% & \multicolumn{1}{c|}{85.3\%} & 100.0\% & \multicolumn{1}{c|}{92.0\%} & 93.3\% & 90.7\% \\
\multicolumn{1}{|l|}{Relevant} & 46.7\% & \multicolumn{1}{c|}{26.7\%} & 100.0\% & \multicolumn{1}{c|}{93.3\%} & 100.0\% & \multicolumn{1}{c|}{93.3\%} & 100.0\% & 97.3\% \\
\multicolumn{1}{|l|}{Correct \& complete} & 33.3\% & \multicolumn{1}{c|}{13.3\%} & 93.3\% & \multicolumn{1}{c|}{82.7\%} & \textbf{100.0\%} & \multicolumn{1}{c|}{\textbf{92.0\%}} & 93.3\% & 88.0\% \\ \hline
\multicolumn{9}{|c|}{\textit{Synth Baseline}} \\ \hline
\multicolumn{1}{|l|}{Concept accurate} & 94.7\% & \multicolumn{1}{c|}{89.5\%} & 100.0\% & \multicolumn{1}{c|}{97.9\%} & 100.0\% & \multicolumn{1}{c|}{100.0\%} & 100.0\% & 100.0\% \\
\multicolumn{1}{|l|}{No inaccuracies} & 94.7\% & \multicolumn{1}{c|}{82.1\%} & 100.0\% & \multicolumn{1}{c|}{96.8\%} & 100.0\% & \multicolumn{1}{c|}{100.0\%} & 100.0\% & 97.9\% \\
\multicolumn{1}{|l|}{Relevant} & 94.7\% & \multicolumn{1}{c|}{90.5\%} & 100.0\% & \multicolumn{1}{c|}{97.9\%} & 100.0\% & \multicolumn{1}{c|}{100.0\%} & 100.0\% & 100.0\% \\
\multicolumn{1}{|l|}{Correct \& complete} & 94.7\% & \multicolumn{1}{c|}{80.0\%} & \textbf{100.0\%} & \multicolumn{1}{c|}{96.8\%} & \textbf{100.0\%} & \multicolumn{1}{c|}{95.8\%} & \textbf{100.0\%} & \textbf{97.9\%} \\ \hline
\multicolumn{9}{|c|}{\textit{Synth RAG}} \\ \hline
\multicolumn{1}{|l|}{Concept accurate} & 100.0\% & \multicolumn{1}{c|}{95.8\%} & 100.0\% & \multicolumn{1}{c|}{97.9\%} & 100.0\% & \multicolumn{1}{c|}{100.0\%} & 100.0\% & 98.9\% \\
\multicolumn{1}{|l|}{No inaccuracies} & 100.0\% & \multicolumn{1}{c|}{81.1\%} & 100.0\% & \multicolumn{1}{c|}{96.8\%} & 100.0\% & \multicolumn{1}{c|}{100.0\%} & 100.0\% & 97.9\% \\
\multicolumn{1}{|l|}{Relevant} & 100.0\% & \multicolumn{1}{c|}{93.7\%} & 100.0\% & \multicolumn{1}{c|}{97.9\%} & 100.0\% & \multicolumn{1}{c|}{100.0\%} & 100.0\% & 98.9\% \\
\multicolumn{1}{|l|}{Correct \& complete} & \textbf{100.0\%} & \multicolumn{1}{c|}{81.1\%} & \textbf{100.0\%} & \multicolumn{1}{c|}{96.8\%} & \textbf{100.0\%} & \multicolumn{1}{c|}{\textbf{100.0\%}} & \textbf{100.0\%} & 97.9\% \\ \hline
\end{tabular}%
\end{table*}

%% file: sec/06Discussion.tex
\section{Discussion}

\subsection{Other Observations}
In the output of all models, we also measured how often a solution is provided. Although this behaviour is not one of the core scoring indicators, it can reflect the model's ability to understand the prompt.
It is worth noting that many models spontaneously provide solution suggestions even when given a system prompt where we explicitly stated "do not give code." 
In the baseline, all models almost completely comply with the prompt, indicating that they can understand and execute the task specification. 
However, after introducing RAG, this indicator dropped significantly on all models, especially in the synthesis-RAG and runtime-RAG settings, where multiple models provided solutions in almost all answers (Close to 100\% of the responses included an attempted "solution").
This phenomenon shows that although RAG can bring information gain and improve diagnostic performance, it may also interfere with the model's compliance with the original task role.
 
\subsection{Threats to Validity}
We adopted a variety of control measures in the experimental design, but there are still the following limitations that may affect the generalization of the results:

\begin{itemize}[leftmargin=*]

\item \textbf{Limited dataset size:} Our dataset contains 34 error samples (19 synthesis, 15 runtime). Although it is carefully constructed to be diverse, it is still not enough to fully cover the HDL error modes in actual production environments.

\item \textbf{Manual annotation:}  All model outputs are manually scored by researchers. Subjectivity and scalability are potential limitations.

\item \textbf{Single prompt structure:} All prompts are single-round dialogues, using a fixed template. In practice, multiple rounds of interaction are possible and would affect overall model performance.

\item \textbf{Frequent model version updates:} Most API LLMs (such as GPT-4o) have iterative changes. The evaluation of this experiment was completed between April and June 2025. Different versions in the future may lead to different results.

\end{itemize}

\subsection{Observations and Takeaways}

This study provides preliminary validation of LLM in HDL debugging tasks and also leads to several broader research directions. 
Many runtime errors (such as timing conflicts, undefined states) are inherently security risks. In the future, LLMs are expected to play an important role in hardware security analysis and verification, explaining potential vulnerabilities in natural language.
Currently, the experiment uses manual evaluation to ensure standard consistency, but this process is time-consuming and has limited scalability. In the future, LLM self-scoring to determine the metrics for answers is also a way forward.
In summary, although current large models have strong capabilities in synthesis errors, there is still great potential in runtime and security-related tasks.

%% file: sec/07Conclusions.tex
\section{Conclusions}
This paper systematically evaluates the performance of large language models (LLMs) in HDL debugging tasks, covering two main categories: synthesis errors and runtime errors. We construct an annotated dataset and evaluate the performance of different models under baseline and RAG-enhanced conditions.
Experimental results show that the current mainstream models have strong capabilities in synthesis error recognition, while the RAG-enhanced mechanism brings significant improvements to most models in runtime errors. GPT-4 and Deepseek perform particularly well in stability and explanation completeness, while GPT-3.5-turbo shows performance degradation under RAG conditions.
In addition, this paper discusses the potential of LLM in hardware security analysis, automatic scoring, and other directions. We believe that this work provides a useful starting point for the future application of large language models to hardware debugging.

\section*{Acknowledgements}
This research is supported in part by the Natural Sciences and Engineering Research Council of Canada (NSERC) [RGPIN-2022-03027]. Cette recherche a été financée en partie par le Conseil de recherches en sciences naturelles et en génie du Canada (CRSNG). 
This research work is partly supported by a gift from Intel Corporation as well as a gift from Google. %

%% file: _main.bbl
\begin{thebibliography}{10}
\providecommand{\url}[1]{#1}
\csname url@samestyle\endcsname
\providecommand{\newblock}{\relax}
\providecommand{\bibinfo}[2]{#2}
\providecommand{\BIBentrySTDinterwordspacing}{\spaceskip=0pt\relax}
\providecommand{\BIBentryALTinterwordstretchfactor}{4}
\providecommand{\BIBentryALTinterwordspacing}{\spaceskip=\fontdimen2\font plus
\BIBentryALTinterwordstretchfactor\fontdimen3\font minus \fontdimen4\font\relax}
\providecommand{\BIBforeignlanguage}[2]{{%
\expandafter\ifx\csname l@#1\endcsname\relax
\typeout{** WARNING: IEEEtran.bst: No hyphenation pattern has been}%
\typeout{** loaded for the language `#1'. Using the pattern for}%
\typeout{** the default language instead.}%
\else
\language=\csname l@#1\endcsname
\fi
#2}}
\providecommand{\BIBdecl}{\relax}
\BIBdecl

\bibitem{dessouky_hardfails_2019}
G.~Dessouky, D.~Gens, P.~Haney, G.~Persyn, A.~Kanuparthi, H.~Khattri, J.~Fung, A.-R. Sadeghi, and J.~Rajendran, ``Hardfails: {Insights} into {Software}-{Exploitable} {Hardware} {Bugs},'' in \emph{Proc. 28th {USENIX} {Conf.} on {Security} {Symp.}}\hskip 1em plus 0.5em minus 0.4em\relax Santa Clara, CA, USA: USENIX Association, 2019, pp. 213--230.

\bibitem{tsai_rtlfixer_2024}
Y.~Tsai, M.~Liu, and H.~Ren, ``{RTLFixer}: {Automatically} {Fixing} {RTL} {Syntax} {Errors} with {Large} {Language} {Model},'' in \emph{Proc. 61st {ACM}/{IEEE} {Design} {Automation} {Conf.}}\hskip 1em plus 0.5em minus 0.4em\relax New York, NY, USA: Association for Computing Machinery, Nov. 2024, pp. 1--6.

\bibitem{liu_invited_2023}
M.~Liu, N.~Pinckney, B.~Khailany, and H.~Ren, ``Invited {Paper}: {VerilogEval}: {Evaluating} {Large} {Language} {Models} for {Verilog} {Code} {Generation},'' in \emph{2023 {IEEE}/{ACM} {Int.} {Conf.} on {Computer} {Aided} {Design} ({ICCAD})}, Oct. 2023, pp. 1--8.

\bibitem{pinckney_revisiting_2025}
N.~Pinckney, C.~Batten, M.~Liu, H.~Ren, and B.~Khailany, ``Revisiting {VerilogEval}: {A} {Year} of {Improvements} in {Large}-{Language} {Models} for {Hardware} {Code} {Generation},'' \emph{ACM Trans. Des. Autom. Electron. Syst.}, Feb. 2025.

\bibitem{blocklove_automatically_2025}
J.~Blocklove, S.~Thakur, B.~Tan, H.~Pearce, S.~Garg, and R.~Karri, ``Automatically {Improving} {LLM}-based {Verilog} {Generation} using {EDA} {Tool} {Feedback},'' \emph{ACM Trans. Des. Autom. Electron. Syst.}, Mar. 2025.

\bibitem{fu_gpt4aigchip_2023}
Y.~Fu, Y.~Zhang, Z.~Yu, S.~Li, Z.~Ye, C.~Li, C.~Wan, and Y.~C. Lin, ``{GPT4AIGChip}: {Towards} {Next}-{Generation} {AI} {Accelerator} {Design} {Automation} via {Large} {Language} {Models},'' in \emph{2023 {IEEE}/{ACM} {Int.} {Conf.} on {Computer} {Aided} {Design} ({ICCAD})}, Oct. 2023, pp. 1--9.

\bibitem{blocklove_chip-chat_2023}
J.~Blocklove, S.~Garg, R.~Karri, and H.~Pearce, ``Chip-{Chat}: {Challenges} and {Opportunities} in {Conversational} {Hardware} {Design},'' in \emph{2023 {ACM}/{IEEE} 5th {Workshop} on {Machine} {Learning} for {CAD} ({MLCAD})}, Sep. 2023, pp. 1--6.

\bibitem{qiu_llm-aided_2024}
S.~Qiu, B.~Tan, and H.~Pearce, ``{LLM}-aided explanations of {EDA} synthesis errors,'' in \emph{2024 {IEEE} {LLM} {Aided} {Design} {Workshop} ({LAD})}, Jun. 2024, pp. 1--6.

\bibitem{pearce_dave_2020}
H.~Pearce, B.~Tan, and R.~Karri, ``\BIBforeignlanguage{en}{{DAVE}: {Deriving} {Automatically} {Verilog} from {English}},'' in \emph{\BIBforeignlanguage{en}{Proc. 2020 {ACM}/{IEEE} {Workshop} on {Machine} {Learning} for {CAD}}}.\hskip 1em plus 0.5em minus 0.4em\relax Virtual Event Iceland: ACM, Nov. 2020, pp. 27--32.

\bibitem{maddala_laag-rv_2024}
K.~Maddala, B.~Mali, and C.~Karfa, ``{LAAG}-{RV}: {LLM} {Assisted} {Assertion} {Generation} for {RTL} {Design} {Verification},'' in \emph{2024 {IEEE} 8th {Int.} {Test} {Conf.} {India} ({ITC} {India})}, Jul. 2024, pp. 1--6.

\bibitem{qiu_autobench_2024}
R.~Qiu, G.~L. Zhang, R.~Drechsler, U.~Schlichtmann, and B.~Li, ``{AutoBench}: {Automatic} {Testbench} {Generation} and {Evaluation} {Using} {LLMs} for {HDL} {Design},'' in \emph{Proc. 2024 {ACM}/{IEEE} {International} {Symposium} on {Machine} {Learning} for {CAD}}, ser. {MLCAD} '24.\hskip 1em plus 0.5em minus 0.4em\relax New York, NY, USA: Association for Computing Machinery, Sep. 2024, pp. 1--10.

\bibitem{kande_security_2024}
R.~Kande, H.~Pearce, B.~Tan, B.~Dolan-Gavitt, S.~Thakur, R.~Karri, and J.~Rajendran, ``({Security}) {Assertions} by {Large} {Language} {Models},'' \emph{{IEEE} Trans. Inf. Forensics Security}, pp. 1--1, 2024.

\bibitem{ahmad_fixing_2023}
B.~Ahmad, S.~Thakur, B.~Tan, R.~Karri, and H.~Pearce, ``On hardware security bug code fixes by prompting large language models,'' \emph{{IEEE} Trans. Inf. Forensics Security}, vol.~19, pp. 4043--4057, 2024.

\bibitem{ahmad_flag_2025}
B.~Ahmad, J.~Ah-kiow, B.~Tan, R.~Karri, and H.~Pearce, ``{FLAG}: {Finding} {Line} {Anomalies} (in {RTL} code) with {Generative} {AI},'' \emph{ACM Trans. Des. Autom. Electron. Syst.}, May 2025.

\bibitem{ferreira_mello_technology_2024}
R.~Ferreira~Mello, N.~Rummel, I.~Jivet, G.~Pishtari, and J.~A. Ruipérez~Valiente, Eds., \emph{\BIBforeignlanguage{en}{Technology {Enhanced} {Learning} for {Inclusive} and {Equitable} {Quality} {Education}: 19th {European} {Conf.} on {Technology} {Enhanced} {Learning}, {EC}-{TEL} 2024, {Krems}, {Austria}, {September} 16–20, 2024, {Proceedings}, {Part} {I}}}.\hskip 1em plus 0.5em minus 0.4em\relax Cham: Springer Nature Switzerland, 2024, vol. 15159.

\bibitem{taylor_dcc_2024}
A.~Taylor, A.~Vassar, J.~Renzella, and H.~Pearce, ``dcc --help: {Transforming} the {Role} of the {Compiler} by {Generating} {Context}-{Aware} {Error} {Explanations} with {Large} {Language} {Models},'' in \emph{Proc. 55th {ACM} {Technical} {Symp.} on {Computer} {Science} {Education} {V}. 1}.\hskip 1em plus 0.5em minus 0.4em\relax New York, NY, USA: Association for Computing Machinery, Mar. 2024, pp. 1314--1320.

\bibitem{karvelas_effects_2020}
I.~Karvelas, A.~Li, and B.~A. Becker, ``The {Effects} of {Compilation} {Mechanisms} and {Error} {Message} {Presentation} on {Novice} {Programmer} {Behavior},'' in \emph{Proc. 51st {ACM} {Technical} {Symp.} on {Computer} {Science} {Education}}.\hskip 1em plus 0.5em minus 0.4em\relax New York, NY, USA: Association for Computing Machinery, Feb. 2020, pp. 759--765.

\bibitem{chen2021evaluating}
M.~Chen, J.~Tworek, H.~Jun, Q.~Yuan, H.~P. D.~O. Pinto, J.~Kaplan, H.~Edwards, Y.~Burda, N.~Joseph, G.~Brockman \emph{et~al.}, ``Evaluating large language models trained on code,'' \emph{arXiv preprint arXiv:2107.03374}, 2021.

\end{thebibliography}
